\documentclass[journal,12pt,onecolumn,draftclsnofoot]{IEEEtran}
\usepackage{float}
\usepackage{cite}
\usepackage{url}
\usepackage{tikz}
\usepackage[ruled,lined,boxed]{algorithm2e}
\usepackage{amsfonts}
\usepackage{amssymb}
\usepackage{amsthm}
\usepackage[cmex10]{amsmath}
\usepackage{booktabs}
\usepackage{hyperref}
\usepackage{mathtools}
\usepackage{microtype}
\usepackage{nicefrac}
\usepackage{url}
\usepackage{xcolor}
\usepackage{cleveref}  %

\newcommand\given{\,\vert\,}  %

\DeclarePairedDelimiterX{\divergence}[2]{(}{)}{#1\,\delimsize\|\,#2}
\newcommand{\kl}{D_\mathrm{KL}\divergence}

\newcommand{\E}{\mathbb{E}} %

\DeclarePairedDelimiter\abs{\lvert}{\rvert}  %
\newcommand\jurl[1]{\href{https://#1}{\nolinkurl{#1}}}

\DeclareMathOperator{\X}{\mathcal{X}}
\DeclareMathOperator{\A}{\mathcal{A}}

\DeclareMathOperator{\BigO}{\mathcal{O}}

\newcommand{\M}{\mathcal{M}}
\newcommand{\B}{\mathcal{B}}
\newcommand{\1}{\mathbf{1}}

\newcommand{\Y}{\mathcal{Y}}
\newcommand{\J}{\mathcal{J}}

\newcommand{\encode}{\mathtt{encode}}
\newcommand{\decode}{\mathtt{decode}}
\newcommand{\flookup}{\mathtt{forward\_lookup}}
\newcommand{\rlookup}{\mathtt{reverse\_lookup}}
\newcommand{\bstinsert}{\mathtt{insert}}
\newcommand{\bstremove}{\mathtt{remove}}

\newcommand{\rebuttal}[1]{#1}
\newcommand{\rebuttaltwo}[1]{{#1}}
\newcommand{\changed}[1]{{#1}}
\newcommand{\multiset}[1]{\rebuttal{\{\!\!\{#1\}\!\!\}}}

\begin{document}
\title{Compressing Multisets with Large Alphabets using Bits-Back Coding}
\author{%
  Daniel Severo$^{123}$
  \and James Townsend$^{4}$
  \and Ashish Khisti$^{2}$
  \and Alireza Makhzani$^{23}$
  \and Karen~Ullrich$^1$
    {\begin{minipage}{\linewidth}\begin{center}
    \renewcommand{\arraystretch}{0.6}
    \begin{tabular}{cc}
                                      &                           \\
    $^{1}$Meta AI               &$^{2}$University of Toronto  \\
    \normalsize{dsevero@fb.com}              &\normalsize{d.severo@mail.utoronto.ca}\\
    \normalsize{karenu@fb.com}               &\normalsize{akhisti@ece.utoronto.ca}\\
                                      &                             \\
    $^3$Vector Institute for AI       &$^4$University of Amsterdam\\
    \normalsize{makhzani@vectorinstitute.ai} &\normalsize{j.h.n.townsend@uva.nl}
    \end{tabular}
    \end{center}\end{minipage}}
}

\maketitle

\begin{abstract}
Current methods which compress multisets at an optimal rate have computational complexity that scales linearly with alphabet size, making them too slow to be practical in many real-world settings.
We show how to convert a compression algorithm for sequences into one for multisets, in exchange for an additional complexity term that is quasi-linear in sequence length.
\changed{This allows us to compress multisets of exchangeable symbols at an optimal rate, with computational complexity decoupled from the alphabet size.}
The key insight is to avoid encoding the multiset directly, and instead compress a proxy sequence, using a technique called `bits-back coding'.
We demonstrate the method experimentally on tasks which are intractable with previous optimal-rate methods: compression of multisets of images and JavaScript Object Notation (JSON) files.
Code for our experiments is available at \url{https://github.com/facebookresearch/multiset-compression}.
\end{abstract}

\section{Introduction}
\changed{
Lossless compression algorithms typically preserve the ordering of compressed symbols in the input sequence.
However, there are data types where order is not meaningful, such as collections of files, rows in a database, nodes in a graph, and, notably, datasets in machine learning applications.
Formally, these may be expressed as a mathematical object known as a \emph{multiset}: a generalization of a set that allows for repetition of elements.

Compressing a multiset with an arithmetic coder is possible if we somehow order its elements and communicate the corresponding ordered sequence. However, unless the order information is somehow removed during the encoding process, this procedure will be sub-optimal, because the order contains information and therefore more bits are used to represent the source than are truly necessary.
}

The fundamental limits of multiset compression are well understood and were first investigated by \cite{varshney2006}. The information content of a non-trivial multiset is strictly less than that of a sequence with the same elements, by the number of bits required to represent an ordering, or permutation, of the elements. However, previous optimal-rate algorithms for multiset compression have computation time which scales linearly with the size of the alphabet from which elements are drawn \cite{Steinruecken2016-oy}.

\changed{In this work\footnote{A preliminary version of this work has been presented at a workshop \cite{severo2021your} and conference \cite{severo2021compressing}.} we show how to compress multisets of statistically exchangeable symbols at an optimal rate, with computational complexity independent of the alphabet size, by converting an existing algorithm for sequences into one for multisets. This enables us to compress fixed-size multisets of independent and identically distributed (i.i.d.) symbols\footnote{\rebuttal{Any i.i.d. sequence is also necessarily exchangeable.}} with arbitrarily large alphabets, including multisets of images, where the alphabet size scales exponentially with the number of pixels; and of strings, where it scales exponentially with string length.\footnote{Note we are using `symbol' as shorthand for `multiset element'. In this work such an element may be an entire image or string of characters.}}

The key insight is to avoid encoding the multiset directly, and instead encode a random sequence containing the same elements as the multiset, in $\BigO(n)$ steps, where $n$ is the sequence and multiset size.
This can be done using a technique known as `free energy' or `bits-back' coding \cite{frey1996free, frey1997, townsend2019}, where encoding and decoding operations are interleaved during compression and decompression.
During compression, symbols are sampled without replacement from the multiset and encoded sequentially in the order they are sampled. Sampling is done using a decoder, with the already compressed information used as the random seed for the sample.
This procedure is invertible, because the bits used as the random seed can be losslessly recovered during decompression, using an encoder. The sampling step consumes bits, reducing the message length by exactly the number of bits needed to represent a permutation. Expected and worst-case time complexities of the method are shown in \Cref{tab:complexity}.

\rebuttaltwo{
Our method is attractive in settings where the alphabet size is large, and the symbol distribution assigns most probability mass to a sparse subset of the alphabet.
}
We anticipate that databases and other unordered structured data, in formats like JavaScript Object Notation (JSON), might be realistic use-cases, while for multisets of large objects, such as images or video files, the savings are marginal. We investigate these settings experimentally in \Cref{sec:experiments}.

\changed{
Although our method works for any source of exchangeable symbols, the following sections assumes symbols are i.i.d. to simplify the exposition.}
In \Cref{sec:notation} we establish the notation used throughout this paper.
The problem setting is formally defined in \Cref{sec:problem}.
We give an extensive overview of related work in \Cref{sec:related}.
In this work we are concerned with both the rate and computational complexity of the compression scheme, hence we discuss the computational complexity of entropy coding in \Cref{sec:aec-complexity}.
Our method requires access to a stack-like entropy coder such as \emph{asymmetric numeral systems} (ANS) \cite{duda2009}, which we discuss in \Cref{sec:ans}.
The method itself is then described and analyzed in \Cref{sec:method}\changed{, assuming i.i.d.\ symbols. \Cref{sec:exchangeability} extends and discusses the method for when symbols are exchangeable.}
Finally we showcase our experiments in \Cref{sec:experiments}, and conclude the paper by discussing directions for future work.

\section{Notation}\label{sec:notation}
Random variables are represented by capital letters such as $X,Y,Z$, while their instances are lower case $x,y,z$.
The alphabet of a random variable $X$ is denoted with $\mathcal{X}$, the calligraphic version of the same symbol.
A sequence $X_1, \dots, X_n$ of size $n$ is abbreviated as $X^n$.

All distributions $P_{(\cdot)}$ are discrete and will be sub-indexed by their respective random variables, such as $P_X$ and $P_{X\given Y}$.

Exceptionally, we will refer to both the instance as well as the random variable of a multiset by $\M$, possibly with a sub-index $\M_i$.
The alphabet of $\M$ will not be explicitly referred to, which should avoid any confusion. The number of elements contained in a multiset $\M$, including repetitions, is sometimes referred to as the ``size of $\M$" and is denoted by $\abs{\M}$.
When the elements of a multiset $\M$ are the same as those in a sequence $X^n$, we write $\M = \{X_1, \dots, X_n\} = \multiset{X^n}$.

The number of elements in $\M = \{x_1, \dots, x_n\}$ that are equal to some $x \in \X$ is denoted as $\M(x) = \sum_{i=1}^n \1\{x = x_i\}$.

\rebuttal{Given a multiset $\M$ and an element $x$, 
the multiset $\M\ –\ \{x\}$ has the same elements as $\M$ but with the occurrence count of element $x$ decreased by $1$.
Given two multisets $\M$ and $\M'$, we define $\M - \M'$ to be the resulting multiset after decreasing the occurrences of elements in $\M$ by their occurrence counts in $\M'$.
Occurrence counts cannot be negative, hence $\{x\} - \{x, x\}$ results in the empty multiset $\emptyset$.
Similarly, $\M + \M'$ denotes the additive union of two multisets.

The concept of exchangeability used throughout the text is that of finite exchangeability.
In particular, we say a finite sequence of random variables $X^n$ is exchangeable if for all permutations $\sigma \colon \{1, \dots, n\} \rightarrow \{1, \dots, n\}$ and all \(x_1,\ldots,x_n\); the following holds:
\begin{equation}
    P_{X^n}(x_1, \dots, x_n) = P_{X^n}(x_{\sigma(1)}, \dots, x_{\sigma(n)}).
\end{equation}
}
Logarithms are always base $2$.

\section{Problem Setting}\label{sec:problem}
Given a sequence of i.i.d.\ discrete random variables $X^n = (X_1, \dots, X_n)$ with symbol alphabet $\X$, the goal is to losslessly encode the \emph{multiset} $\M = \{X_1, \dots, X_n\} = \multiset{X^n}$ at rate
\begin{equation}\label{eq:ms-entropy}
    H(\M) = H(X^n) - H(X^n \given \M) = I(\M; X^n),
\end{equation}
which follows from $H(\M\given X^n) = 0$. This problem can equivalently be seen as that of encoding the sequence $X^n$ with complete disregard to the order between symbols $X_i$.

The term $H(X^n\given\M)$ can be interpreted as the average number of bits required to order the elements in $\M$ to create a sequence $X^n$. For this reason this term is sometimes referred to as the \emph{order information} \cite{varshney2006}.

For i.i.d.\ symbols the probability of the multiset depends only on the symbols $\multiset{X^n}=\multiset{x^n}$, irrespective of their relative ordering. We express this as
\begin{align}
    P_\M(\M) &= M P_{X^n}(x^n),
\end{align}
where the constant $M$ is known as the \emph{multinomial coefficient} of $\M$. This coefficient is equal to the number of unique permutations of $X^n$
\begin{align}
    M = \abs{\left\{x^n \in \X^n: \multiset{x^n} = \M \right\}}
    = \frac{n!}{\prod_{x\in\X }\M(x)!} \leq n!.
\end{align}

We are concerned with both the rate and computational complexity of encoding and decoding. Although the method we present is applicable to any alphabet $\X$ and multiset size $\abs{\M}=n$, we are mainly interested in sources with large alphabets $\abs{\X} \gg n$. For example, if $X_i$ are images then $\abs{\X}$ will grow exponentially with the number of pixels.

\section{Related Work}\label{sec:related}
To the best of our knowledge, there are no previous works that present a method which is both computationally feasible and rate-optimal for compressing multisets of i.i.d.\ symbols with large alphabets.

The fundamental limits on lossless compression of multisets were investigated
in \cite{varshney2006}. For finite alphabets, a sequence can be decomposed into
a multiset $\M$ (also known as a \emph{type}) and a permutation conditioned on
the multiset. With this, the authors are able to prove a combinatorial bound on
the entropy: $H(\M) \leq \BigO(|\X|\log|\M|)$. However, a practical algorithm
for achieving this bound is not presented. 

A rate-optimal algorithm for compressing multisets of i.i.d\ symbols was put
forth in \cite{Steinruecken2016-oy}. There, the information content of the
multiset is decomposed recursively over the entire alphabet, allowing it to be
interfaced to an arithmetic coder. Alternatively, this can also be achieved
through Poissonification, even for arbitrary-sized multisets
\cite{steinruecken2014b, yang2014compression, yang2017minimax}. Although
rate-optimal, the computational complexity of these methods scales linearly
with alphabet size.

Another approach \cite{gripon2012compressing, Steinruecken2014-zs,
reznik2011coding} is to convert multisets of sequences (e.g., cryptographic
hashes) to an order-invariant data structure, such as a tree, that is
losslessly compressible. These methods can compress arbitrary multisets of
i.i.d.\ symbols by first encoding each symbol with a prefix-free code and
constructing the order-invariant data-structure from the multiset of resulting
code-words. If encoding each symbol individually is optimal (e.g., Huffman
coding for dyadic sources), then the method presented in
\cite{Steinruecken2014-zs} is optimal and has the same complexity as our
method. However, this is not generally the case, and the overhead of
compressing each symbol with a prefix-free code may dominate. Our method can be
seen as a generalization of \cite{Steinruecken2014-zs} that allows symbols to
be compressed in sequence, effectively removing the overhead.

Compressing a multiset is equivalent to compressing a \emph{sequence with known order} \cite{steinruecken2014b}, as both objects have equal information content.
The transmitter and receiver can agree that sequences will be sorted before encoding, which removes the order information.
An optimal coding scheme could compress sorted symbols sequentially with an adaptive symbol distribution that accounts for sorting.
However, like the method in \cite{Steinruecken2016-oy}, this approach does not scale to multisets with large alphabet \changed{due to the high complexity cost of adapting the symbol distribution at each encoding step \cite{steinruecken2014b}.}

An alternative to removing the order information is to transform the sequence in a way that aids compression. 
A common choice of transformations is to sort and apply run-length encoding (RLE) \cite{robinson1967results}.
The output is a set of symbol-frequency pairs, known as \emph{run-length symbols}, which has the same information content as the multiset and an alphabet of size $\abs{\X}n$.
Compressing the set of run-length symbols as a sequence with known order is computationally intensive (as previously discussed), but can be done efficiently with our method.
However, the entropy of any transformation is lower bounded by that of the multiset, as we require that the multiset be recoverable.
Therefore, these techniques can not improve the average compression rate, although they might outperform our method for carefully chosen examples. 

A related but more complex setting is \emph{lossless dataset compression},
where the order between correlated examples in a dataset is irrelevant. A
sub-optimal two-step algorithm that first re-orders the dataset based on
similarity between examples, followed by predictive coding is proposed in
\cite{Barowsky2021-wg}. A neural network is overfitted on the dataset to
predict pixel values autoregressively based on the image to which the pixel
belongs, as well as previously seen images. This setting is equivalent to
sub-optimally compressing a multiset of correlated symbols, while our method
optimally compresses multisets of i.i.d.\ symbols.

\section{Complexity of adaptive entropy coding}\label{sec:aec-complexity}
This work deals with the computational complexity of rate-optimal entropy coding.
In this section, we discuss some aspects of the source distribution that affect the computational complexity of encoding/decoding in adaptive entropy coding (AEC).

The objective of AEC is to losslessly encode/decode an arbitrary random sequence $X^n = (X_1, \dots, X_n)$ at the rate equal to its \emph{entropy rate}, $\frac{1}{n}H(X^n)$.
We assume, without loss of generality, that all $X_i$ share a common alphabet $\X$.
At the $i$-th step, the symbol $X_i=x_i$ is encoded/decoded using the conditional distribution $P_i = P_{X_i\given X^{i-1}}$, where $X^{i-1}=x^{i-1}$ is the sequence of previously observed symbols or \emph{context}.

To achieve this one can employ range-based entropy coders such as Arithmetic Coding and Asymmetric Numeral Systems.
Encoding and decoding require computing the \emph{range} $R_i(x_i) = \left[F_i(x_i), F_i(x_i) + P_i(x_i)\right)$, where $F_{(\cdot)}$ represents the cumulative distribution of the corresponding $P_{(\cdot)}$.

The computational complexity of computing the range is critical to the overall complexity of encoding and decoding, and can be done adaptively as the context changes.
For example, consider the case of encoding $x_i$ with distribution $P_i(x_i)=f_i(x_i)/(i-1)$ where $f_i(x_i) = \sum_{j=1}^{i-1}\1\left\{x_i = x_j\right\}$ is the frequency count of $x_i$ in the context.
Storing the frequency counts $\{f_i(x): x \in \X\}$ in an array of size $\abs{\X}$ allows $\BigO(1)$ worst-case lookups of $P_i(x_i)$ and $\BigO(n)$ for $F_i(x_i) = \sum_{x \in \X: x < x_i} f_i(x)$ via a cumulative sum, totalling $\BigO(n)$ to compute the range.
In preparation for the encoding/decoding of the next symbol $x_{i+1}$, the array can be updated by incrementing the count $f_i(x_i)$ to $f_i(x_i)+1$, also in $\BigO(1)$.

In the extreme case where symbols are i.i.d., no adaptation of the datastructure is necessary as $P_i = P_1$ for all $i$.
The overall complexity of encoding/decoding will depend mostly on the cost of computing the range.
We sometimes refer to this setting as \emph{static entropy coding}.

The complexity of adapting the datastructure to the next symbol as well as that of computing the range can drastically change depending on the choice of datastructure.
For example, using a binary search tree (BST) instead of an array decreases the worst-case complexity of computing the range to $\BigO(\log n)$, while increasing that of adapting to the same cost.

\section{Asymmetric Numeral Systems}\label{sec:ans}
The multiset compression method in this paper depends on asymmetric numeral systems (ANS) \cite{duda2009}, \rebuttal{a family of} last-in-first-out, or stack-like entropy coders which we describe in this section.
\rebuttal{In this section, we describe an idealized version of ANS that uses a single unbounded integer as the compressed state. Specific variants of ANS, such as rANS that uses a bounded integer together with a bitstream \cite{duda2009}, are captured by this description.}
A slightly more detailed description of ANS is provided in \cite{townsend2021lossless}.

ANS encoding moves information into a (large) natural number, which we will denote \rebuttaltwo{\(s \in \mathbb{N}\)}.\footnote{\rebuttal{
Our description extends to rANS/tANS by mapping the integer state to the higher order bits of $s$ and the contents of the bitstream to the lower-order bits.
\rebuttaltwo{
If the integer state of rANS/tANS is denoted $t$, and the bits in the bitstream are denoted $b_1,\ldots,b_n$, then $s = 2^n t + \sum_{i=1}^n 2^{i-1}b_i$.}
}}
The outputted bitstream is the binary representation of $s$, which is approximately $\log s$ bits in size.
For a symbol \(X=x\) from alphabet $\X$ and probability mass function \(P_X\), the encode and decode functions form an inverse pair:
\begin{align}
  \encode&\colon (s, x)\mapsto s'\\
  \decode&\colon s'\mapsto (s, x),
\end{align}
where
\begin{equation}\label{eq:ans-approx}
  \log s'\approx \log s + \log 1/P_X(x).
\end{equation}
Encoding a sequence of i.i.d.\ symbols $X^n=x^n$ with ANS proceeds by initializing the state $s_0$, followed by successively applying $s_i = \encode(s_{i-1}, x_i)$ with $P_X$.
The final ANS state $s_n$ holds the information of the entire sequence, which will be approximately $\log s_n \approx \sum_{i=1}^n \log 1 / P_X(x_i)$ bits in size. 

Decoding with ANS successively applies $(s_{i-1}, x_i) = \decode(s_i)$ until the initial state $s_0$ is recovered.
The first decoded symbols will be $x_n$, which was the last one encoded.
For this reason, the ANS state is sometimes referred to as the ``ANS stack".

The inaccuracy of \cref{eq:ans-approx} is typically small, and in a standard
implementation can be bounded by \(\epsilon =2.2\times10^{-5}\) bits per
operation, equivalent to one bit of redundancy for every 45,000 operations
\cite{townsend2020}.
In addition to the small per-symbol redundancy, there is also
a one-time redundancy incurred when initializing and terminating encoding, more
details are given in the \cite{townsend2021lossless}.

ANS requires that the distribution $P_X$ be quantized. The quantized distribution must be specified by a precision parameter $N \in \mathbb{N}$
together with two lookup functions:
\begin{align}
  \label{eq:lookup}
  \flookup &\colon x\mapsto (c_x, p_x)\nonumber\\
  \rlookup &\colon i\mapsto (x, c_x, p_x),
\end{align}
used respectively by $\encode$ and $\decode$.
The integers \(c_x\) and \(p_x\) can be thought of as a quantized cumulative
distribution function (CDF) and probability mass, with the finite set \(\{0,
\ldots, N-1\}\) modeling the real interval \([0, 1]\). They must satisfy
\begin{equation}\label{eq:quantized-pmf}
  \frac{p_x}{N} = P_X(x) \qquad 
  c_x = \sum_{y\in\X:y < x} p_y,
\end{equation}
for all $x \in \X$.
The $\rlookup$ has input \(i\in \{0,\ldots,N-1\}\) and must
return the unique triple \((x, c_x, p_x)\) such that \(i\in[c_x,c_x + p_x)\).

It is possible to mix ANS with any prefix-free code by inserting the code-words into lower-order bits of the ANS state \rebuttal{(see page 6 of \cite{giesen2014} for more details on this technique)}.
We use this technique in our experiments to encode a multiset of images compressed with the lossy WebP codec.

Key to our method is the observation that ANS can be used for sampling by randomly initializing the state to a \rebuttal{non-negative} integer value $s'$, followed by applying $(s, x) = \decode(s')$ with the distribution $P_X$ one wishes to sample from.
\rebuttal{
ANS decode performs integer arithmetic on $s'$ to compute $s$ and $x$ (see equation 2.26 of \cite{townsend2021lossless}). The construction of ANS guarantees that \emph{any non-negative integer} $s'$ is a valid input to $\decode$ (see Figures 2.2 and 2.3 of \cite{townsend2021lossless}).
Furthermore, any $s'$ that is the output of an $\encode$ step is guaranteed to be a valid input to future $\decode$ operations, as the output of $\encode$ is always a non-negative integer.
This implies the state need not have been randomly initialized, but may instead have been constructed from a sequence of $\encode$ steps from a previous encoding task, as the origin of the state $s'$ is irrelevant.}
This allows the sampling distribution $P_X$ to differ from the distribution used for the previous task.
The ANS state is thus used as a \emph{random seed}, and because decoding/sampling $X=x$ removes approximately $-\log P_X(x)$ bits from the state, the random seed is slowly consumed as symbols are sampled.
The operation is invertible in the sense that the random seed can be recovered by performing $\encode$ with the sampled symbol, i.e.,\ $\encode(\decode(s')) = s'$.

The idea of using ANS for sampling comes from recent work in the machine learning community, where it has been referred to as \emph{bits-back coding with ANS} (BB-ANS).
The idea can be traced back to \cite{frey1997}, 12 years prior to the invention of ANS, which proposed to achieve a similar effect, albeit with a sub-optimal compression rate, using Huffman coding.
The critical insight, that mixing encoding and decoding is significantly more straightforward, and can achieve an optimal compression rate with ANS, was made by \cite{townsend2019}, who demonstrated it by implementing compression of images with \emph{latent variable models}.
This has been followed by a series of works proposing elaborations on the original idea for specific classes of latent variable models \cite{Kingma2019-vx,townsend2020a,townsend2021a} and for improving the compression rate by using Monte Carlo methods \cite{ruan2021}.
The distribution we use over multisets can be viewed as a latent variable model where the multiset is observed and the ordering is latent.

\section{Bits-back coding with ANS (BB-ANS)}\label{sec:bbans}
BB-ANS can be used to significantly reduce the computational complexity of adaptive entropy coding (AEC, see \Cref{sec:aec-complexity}) without changing the achieved rate. Our method uses BB-ANS to compress a multiset via a proxy sequence of i.i.d.\ symbols. In this section, we describe BB-ANS in a more general form.

\rebuttaltwo{Let $X, Y$ represent random variables where $X = f(Y   )$ is a deterministic function of $Y$ with $f\colon \Y \mapsto \X$.}
Our objective is to encode $X$ at rate $H(X)$, which is assumed to require significantly more \rebuttal{computational resources} than encoding $Y$ at rate $H(Y)$ (see \Cref{sec:aec-complexity}).
Directly encoding $Y$ as a stand-in for $X$ would reduce the required \rebuttal{computational resources}, but potentially yields a larger rate $H(Y) \geq H(X)$.
\rebuttal{Instead, our strategy will be to construct a conditional distribution $P_{Y \given X}$ that maps $X$ to a random $Y$ via sampling.}

BB-ANS requires access to an ANS stack $s_0$, which may be randomly initialized or re-purposed from a previous compression task. To encode a symbol $X=x$, BB-ANS first maps $X=x$ to a random value $Y=y$ by decoding from the ANS stack with the corresponding code from the family (see \Cref{sec:ans})
\begin{equation}
   (s_1, y) = \decode(s_0) \text{ with } P_{Y\given X}(\cdot\given x).
\end{equation}
The decoded value is then encoded using the code for $Y=y$
\begin{equation}
       s_2 = \encode(y, s_1) \text{ with } P_{Y}(y).
\end{equation}
Decoding reduces the size of the ANS stack, while encoding increases it. Overall, encoding $X=x$ increases the stack on average by $-H(Y\given X) + H(Y) = H(X)$.
The final stack $s_2$ is then transmitted to the decoder.

Decoding proceeds in reverse order. First, $y$ and $x=f(y)$ are decoded from the ANS stack $s_2$
\begin{align}
   (s_1, y) &= \decode(s_2) \text{ with } P_{Y}(\cdot)\\
        x &= f(y),
\end{align}
and then the initial ANS stack $s_0$ can be recovered via an encode step with $y$
\begin{align}
       s_0 = \encode(y, s_1) \text{ with } P_{Y\given X}(y\given x).
\end{align}
Restoring $s_0$ is necessary if one wishes to apply this procedure sequentially, as we discuss next.

BB-ANS applies this procedure to a collection $X_1, \dots, X_n$ on an initially empty ANS stack, where $X_i \sim X$.
At each step, the ANS stack increases by $H(X_i) = H(X)$ on average.
The very first step would be to decode $Y_1\given X_1$, which is not possible as the ANS stack is empty.
This step is therefore skipped and $H(Y_1\given X_1)$ is considered to be a one-time overhead, which is referred to as the \emph{initial bits} in the BB-ANS literature. Applying BB-ANS to the sequence $X^n$ amortizes the initial bits overhead, and the total rate is thus
\begin{equation}
    \lim_{n \rightarrow \infty} \frac{1}{n}\left( H(Y_1) 
    + \sum_{i=2}^n \left( H(Y_i) - H(Y_i \given X_i)\right) \right) = H(X).
\end{equation}

The computational complexity of BB-ANS is that of decoding $Y\given X$ and encoding $Y$, which may require significantly less \rebuttal{computational resources} than encoding $X$ directly.
In \Cref{sec:method} we show how the careful design of a family of codes for $Y\given X$ can lead to a significant decrease in complexity for encoding and decoding multisets of i.i.d.\ symbols.

\rebuttal{\section{Bits-back Coding for Multiset Compression}}\label{sec:method}
In this section we explain our method, shown in \cref{alg:encode-decode}, and provide proof of the achievable rate $R=H(\M)$.
\footnote{\rebuttal{A video describing our method can be found at \url{https://youtu.be/Gwf9_t-JjsQ?t=758}}} \rebuttal{ A detailed illustration of our method compressing a small multiset is available in Appendix \Cref{sec:method-viz}.}

Achieving the entropy of the multiset $\M$ requires amortizing over a collection of multisets $\M^n$ due to the initial bits overhead.
While amortizing is unavoidable, we show it is still possible to compress a single multiset (i.e., one-shot setting) at a rate equal to its entropy.
The key insight is that we can decompose the task of compressing a single multiset into that of compressing a collection of multisets, which can be used to amortize the initial bits overhead of BB-ANS, while still achieving the entropy.

As in \Cref{sec:bbans}, compressing with the distribution of the source $P_{\M^n}$ is assumed to be prohibitively expensive computationally. We therefore use BB-ANS to map $\M^n$ to a proxy sequence $\tilde{X}^n$ which requires significantly less \rebuttal{computational resources to compress}.

In \Cref{subsec:algorithm} we describe the algorithm and show it achieves rate $R = H(\tilde{X}^n) - H(\tilde{X}^n\given\M)$. In \Cref{subsec:rate} we prove $R=H(\M)$ by interpreting $\tilde{X}^n$ as a sequence of draws from a \emph{multivariate hypergeometric distribution}.

\begin{algorithm*}[t]
\caption{Multiset encode (left) and decode (right) }
\small
\begin{minipage}[t]{0.45\textwidth}
    Given a multiset $\M = \{x_1, \dots, x_n\}$\\
    Initialize an ANS state $s_0$, and $\M_1 = \M$\\
    \For{$i = 1, \dots, n$}{
      $s_i', \tilde{x}_i = \decode(s_{i-1})$ with $P_{X_i\given\M_i}(\cdot\given \M_i)$\\
      $\M_{i+1} = \M_i - \{\tilde{x}_i\}$\\
      $s_i = \encode(s_i', \tilde{x}_i)$ with $P_{X}(\tilde{x}_i)$
    }
    \Return $s_n$
\end{minipage}
\label{alg:encode-decode}
\hfill
\begin{minipage}[t]{0.45\textwidth}
    Given an integer ANS state $s_n$\\
    Initialize $\M_{n+1} = \emptyset$\\
    \For{$i = n, \dots, 1 $}{
        $s_i', \tilde{x}_i = \decode(s_i)$ with $P_{X}(\cdot)$\\
        $\M_i = \M_{i+1} + \{\tilde{x}_i\}$\\
        $s_{i-1} = \encode(s_i', \tilde{x}_i)$ with $P_{X_i\given\M_i}(\tilde{x}_i\given \M_i)$\hfill
    }
    \Return $\M_1 = \M$ 
\end{minipage}
\end{algorithm*}

\subsection{Algorithm}\label{subsec:algorithm}
In this section we describe our algorithm without providing proof of the claimed results, which are given in \Cref{subsec:rate}.

Let $X^n$ represent the i.i.d.\ sequence of source symbols and $\M = \multiset{X^n}$ the corresponding multiset.
Our method uses BB-ANS to map a multiset to a proxy sequence $\tilde{X}^n$ of i.i.d.\ symbols $\tilde{X}_i \sim \tilde{X}$ via a lossless code at rate $H(\tilde{X}^n\given\M)$.
The proxy sequence is then encoded at rate $H(\tilde{X}^n)$ via $n$ sequential encoding steps at rate $H(\tilde{X})$, which requires significantly less \rebuttal{computational resources} than encoding $\M$ directly as static entropy coding can be used.
By construction, our method guarantees that $\tilde{X}^n$ will be a permuted version of $X^n$.

To overcome the initial bits overhead we decompose the task of compressing $\M$ into that of compressing a collection $\M_1, \dots, \M_n$.
Initially, we set $\M_1 = \M$.
At the $i$-th step, a symbol $\tilde{X}_i = \tilde{x}_i$ is sampled without replacement from $\M_i$ with probability proportional to its frequency count $\M_i(\tilde{x}_i)$
\begin{equation}\label{eq:swor-prob}
    P_{\tilde{X}_i\given\M_i}(\tilde{x}_i\given\M_i) 
    = \frac{\M_i(\tilde{x}_i)}{\abs{\M_i}}
    = \frac{\M_i(\tilde{x}_i)}{n -(i-1)},
\end{equation}
yielding the next multiset 
\begin{align}\label{eq:ms-seq-def}
    \M_{i+1} = \M_i - \{\tilde{x}_i\}.
\end{align}
Sampling is performed via a decode on the ANS stack, which contributes $-H(\tilde{X}_i\given\M_i)$ to the rate.
The sampled symbol $\tilde{x}_i$ is then encoded onto the ANS stack with distribution $P_{\tilde{X}}(\tilde{x}_i)$,
increasing it on average by $H(\tilde{X}_i)$. This procedure continues until the multiset is depleted, i.e., $\M_{n+1}=\emptyset$.

As before, the first decode is skipped as the ANS stack is initially empty.
The initial bits overhead is a one-time penalty of $H(\tilde{X}_1\given\M_1)$, which we discuss in detail in \cref{subsec:initbits}. We ignore the initial bits from here on as it does not affect the rate for increasing $n$.

The rate of the proposed scheme, minus the initial bits overhead, is thus 
\begin{align}\label{eq:ms-rate}
    R = \sum_{i=1}^n\left( H(\tilde{X}_i) 
    - H(\tilde{X}_i\given\M_i)\right).
\end{align}

\subsection{Rate}\label{subsec:rate}
We now prove the rate of this scheme \eqref{eq:ms-rate} is equal to the entropy of the multiset $\M = \M_1$ shown in \cref{eq:ms-entropy}.
First, we prove that although $\tilde{X}_i$ are statistically dependent conditioned on $\M$, they are i.i.d. marginally with distribution $P_{\tilde{X}} = P_X$.
This allows static entropy coding of each symbol $\tilde{x}_i$ to be performed with the source distribution $P_X$ and implies the equality \emph{1)} shown below.
Finally, we prove \emph{2)} and conclude from \eqref{eq:ms-rate} that our method achieves $R = H(\M)$.

\

\subsubsection{$\sum_{i=1}^n H(\tilde{X}_i) = H(X^n)$}\label{subsubsec:rate-1}
The sequence $\tilde{X}^n$ is constructed by sampling without replacement from the multiset $\M$. This procedure has been well studied and $\tilde{X}_i$ is known as the $i$-th draw of a multivariate hypergeometric distribution. This process is known to be exchangeable, implying
\begin{align}\label{eq:exch}
    P_{\tilde{X}^n \given\M}(\tilde{x}^n\given \M) 
    & = P_{\tilde{X}^n \given\M}(\sigma(\tilde{x}^n)\given \M) \\
    &= \frac{1}{M}\1\left\{ \multiset{\tilde{x}^n} = \M \right\},
\end{align}
for any permutation $\sigma(\tilde{x}^n) = (\tilde{x}_{\sigma(1)}, \dots, \tilde{x}_{\sigma(n)})$ where $M$ is the multinomial coefficient constructed from the frequency count of symbols in the multiset (see \Cref{sec:problem}). This implies the marginal distribution is
\begin{align}
    P_{\tilde{X}^n}(\tilde{x}^n)
    &= \E\left[ P_{\tilde{X}^n\given\M}(\tilde{x}^n\given\M) \right]\\
    &= \sum_\M \frac{1}{M}\1\left\{\multiset{\tilde{x}^n} = \M\right\} MP_{X^n}(\tilde{x}^n) \\
    &= P_{X^n}(\tilde{x}^n).
\end{align}

\subsubsection{$-\sum_{i=1}^n H(\rebuttal{\tilde{X}_i}\given\M_i) = -H(X^n\given\M)$} 
Given $\M_i, \tilde{X}_i$, the next multiset $\M_{i+1}$ can be constructed from \eqref{eq:ms-seq-def}. Therefore,
\begin{align}
    H(\tilde{X}^n \given \M) 
    &= H(\tilde{X}_1 \given \M_1) + H(\tilde{X}_2 \given \M_1, \tilde{X}_1) + \dots \\
    &= H(\tilde{X}_1 \given \M_1) + H(\tilde{X}_2 \given \M_2) + \dots \\
    &= \sum_{i=1}^n H(\tilde{X}_i\given\M_i).
\end{align}
\rebuttaltwo{
We can show that $H(\tilde{X}^n\given\M) = H(X^n\given\M)$ by noting that $\multiset{\tilde{x}^n} = \multiset{x^n} = \M$ and
\begin{align}
    P_{\tilde{X}^n \given \M}(\tilde{x}^n \given \M)
    = \frac{1}{M} \1\{{\multiset{\tilde{x}^n} = \M}\}
    = P_{X^n \given \M}(\tilde{x}^n \given \M),
\end{align}
which concludes the proof of the rate.

}

\subsection{Efficient Sampling without Replacement}\label{subsec:swor}
For our method to be computationally attractive, we must carefully construct an algorithm that can decode $\tilde{X}_i$ from $\M_i$, as well as adapt to the next symbol $\tilde{X}_{i+1}$ (see \Cref{sec:aec-complexity}).
Decoding the proxy sequence requires sampling without replacement (SWOR) from the multiset.
In this section we show that both SWOR and adapting to the next symbol can be done in sub-linear time for each symbol, or quasi-linear for the entire multiset.
This requires a binary search tree (BST), which we describe in detail next, to store the multisets $\M_i$.
The BST is similar to a Fenwick tree \cite{fenwick1994,moffat1999}, and allows lookup, insertion and removal of elements in time proportional to the depth of the tree. We analyze the overall time complexity of our method in \Cref{subsec:complexity}.

Each encoding/decoding step requires a forward/reverse lookup (see \ref{eq:lookup}) with distribution
$P_{X_i\given\M_i}(\cdot\given \M_i)$. Since we are sampling without replacement, we also
need to perform removals and insertions:
\begin{align}
\bstremove &\colon (\M_i, \tilde{x}_i) \mapsto \M_i - \{\tilde{x}_i\}\\
\bstinsert &\colon (\M_{i+1}, \tilde{x}_i) \mapsto \M_{i+1} + \{\tilde{x}_i\}.
\end{align}
All four operations (forward and reverse lookup, insertion and removal) can be
done in $\BigO(\log m)$ time, where \(m\) is the number of unique elements in
\(\M\) as well as the number of nodes in the BST.

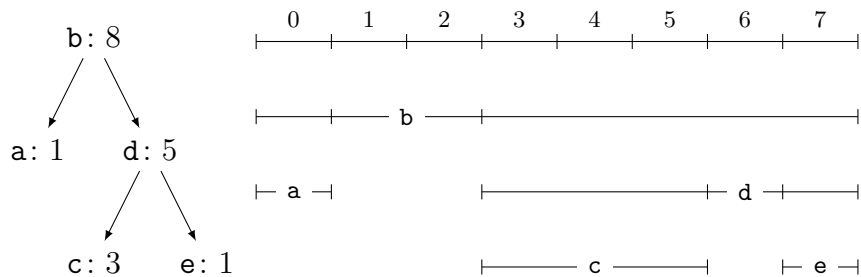
\begin{figure}[ht]
  \centering
    \begin{tikzpicture}[
      edge from parent/.style = {draw,-latex}
      ]
      \node {$\mathtt{b}\colon 8$}
      child {node {$\mathtt{a}\colon 1$}}
      child {node {$\mathtt{d}\colon 5$}
                child {node {$\mathtt{c}\colon 3$}}
                child {node {$\mathtt{e}\colon 1$}}};
    \end{tikzpicture}
    \begin{tikzpicture}
      \draw (0, 0) -- (8, 0);
      \foreach \x in {0,...,8} {
        \draw (\x,.1) -- (\x,-.1) ;
    }
      \foreach \x in {0,...,7} {
        \draw (\x + 0.5, .3) node[fill=white] {\footnotesize\(\x\)};
    }

      \draw (3,   -3)      -- (6, -3);
      \draw (3,   -3 - .1) -- (3, -3 + .1);
      \draw (6,   -3 - .1) -- (6, -3 + .1);
      \draw (4.5, -3) node[fill=white] (c) {\small\(\mathtt{c}\)};

      \draw (7,   -3)      -- (8, -3);
      \draw (7,   -3 - .1) -- (7, -3 + .1);
      \draw (8,   -3 - .1) -- (8, -3 + .1);
      \draw (7.5, -3) node[fill=white] (e) {\small\(\mathtt{e}\)};

      \draw (0,  -2)      -- (1, -2);
      \draw (0,  -2 - .1) -- (0, -2 + .1);
      \draw (1,  -2 - .1) -- (1, -2 + .1);
      \draw (.5, -2) node[fill=white] (a) {\small\(\mathtt{a}\)};

      \draw (3,   -2)      -- (8, -2);
      \draw (3,   -2 - .1) -- (3, -2 + .1);
      \draw (6,   -2 - .1) -- (6, -2 + .1);
      \draw (7,   -2 - .1) -- (7, -2 + .1);
      \draw (8,   -2 - .1) -- (8, -2 + .1);
      \draw (6.5, -2) node[fill=white] (d) {\small\(\mathtt{d}\)};

      \draw (0, -1)      -- (8, -1);
      \draw (0, -1 - .1) -- (0, -1 + .1);
      \draw (1, -1 - .1) -- (1, -1 + .1);
      \draw (3, -1 - .1) -- (3, -1 + .1);
      \draw (8, -1 - .1) -- (8, -1 + .1);
      \draw (2, -1) node[fill=white] (b) {\small\(\mathtt{b}\)};
    \end{tikzpicture}
    \caption{For the multiset $\{\mathtt{a, b, b, c, c, c, d, e}\}$, on the top a schematic representation of the dynamic BST data structure which we use to represent the multiset, and on the bottom the intervals corresponding to each branch of the BST.}\label{fig:bst-1}
\end{figure}

We set the ANS precision parameter \rebuttal{(see \Cref{sec:ans})} to \(N = |\M_i| = |\M| - (i -1) \), with
frequency count $p_x = \M_i(x)$ for each $x \in \M_i$.
At each step $i$, a binary search tree (BST) is used to represent the multiset
of remaining symbols $\M_i$.  The BST stores the frequency and cumulative
counts, $p_x$ and $c_x$, and allows efficient lookups.  At each node is a
unique symbol from \(\M_i\) (arranged in the usual BST manner to enable fast
lookup), as well as a count of the total number of symbols in the entire sub-tree
of which the node is root. \Cref{fig:bst-1} shows an example, for the multiset
$\M_i = \{\mathtt{a, b, b, c, c, c, d, e}\}$.

On the left is the actual BST data-structure, i.e.,\ that which is held in memory
and operated on. The interval diagram on the right visually represents the
local information available from inspection of the counts at a node and its
children. For example, the sub-interval to the right of $\mathtt{b}$ has length
$5$, which is the count at it's child node $\mathtt{d}$.

\changed{
Note that all sub-trees correspond to some sub-multiset $\M' \subseteq \M$.
Under this interpretation, the count at each node represents the size $\abs{\M'}$ of the sub-multiset corresponding to the sub-tree with that particular node as root. The count at the root of the BST ($8$ in \Cref{fig:bst-1}) is equal to the total size $\abs{\M}$ of the multiset.

Below we provide pseudo-code for $\bstinsert, \bstremove, \flookup$ and $\rlookup$. Here the symbol $\B$ is used for the BST that represents $\M$. The BST itself is a 4-tuple $\B = (n, y, \B_L, \B_R)$, where $n=\abs{\M}$ is the size of $\M$, $y$ is the root symbol of $\B$, and $\B_L,\B_R$ are the BSTs with root symbols equal to the left and right children of the root node $y$ of $\B$. We also define $\abs{\B} = \abs{\M} = n$.

\

\begin{algorithm*}[H]
\caption{$\bstinsert$ and $\bstremove$ of symbol $x$ with BST $\B$.}
\small
\begin{minipage}[t]{0.45\textwidth}
  \DontPrintSemicolon
  \SetKwFunction{InsertFunc}{$\bstinsert$}
  \SetKwProg{Fn}{}{:}{}
  \Fn{\InsertFunc{$\B$, $x$}}{
    $(n, y, \B_L, \B_R) = \B$\;
  \uIf{$x < y$}{
    $\B_L = \bstinsert(\B_L, x)$\;
  }
  \uElseIf{$x > y$}{
    $\B_R = \bstinsert(\B_R, x)$\;
  }
  \KwRet $(n+1, y, \B_L, \B_R)$\;
  }
\end{minipage}
\label{alg:insert-remove}
\hfill
\begin{minipage}[t]{0.45\textwidth}
  \DontPrintSemicolon
  \SetKwFunction{RemoveFunc}{$\bstremove$}
  \SetKwProg{Fn}{}{:}{}
  \Fn{\RemoveFunc{$\B$, $x$}}{
    $(n, y, \B_L, \B_R) = \B$\;
  \uIf{$x < y$}{
    $\B_L = \bstremove(\B_L, x)$\;
  }
  \uElseIf{$x > y$}{
    $\B_R = \bstremove(\B_R, x)$\;
  }
  \KwRet $(n-1, y, \B_L, \B_R)$\;
  }
\end{minipage}
\end{algorithm*}
}

\

\changed{
\begin{algorithm*}[H]
\caption{$\flookup$ of symbol $x$, and $\rlookup$ of index $i$, in $\B$.}
\small
\begin{minipage}[t]{0.45\textwidth}
  \DontPrintSemicolon
  \SetKwFunction{FLookupFunc}{$\flookup$}
  \SetKwProg{Fn}{}{:}{}
  \Fn{\FLookupFunc{$\B$, $x$}}{
    $(n, y, \B_L, \B_R) = \B$\;
    $(n_R, n_L) = (\abs{\B_R}, \abs{\B_L})$\;
    $(c_y, p_y) = (n_L, n - (n_L + n_R))$\;
  \uIf{$x = y$}{
    $(c_x, p_x) = (c_y, p_y)$\;
  }
  \uElseIf{$x < y$}{
    $(c_x, p_x) = \flookup(\B_L, x)$\;
  }
  \uElseIf{$x > y$}{
    $(c, p_x) = \flookup(\B_R, x)$\;
    $c_x = c - (n - n_R)$\;
  }
  \KwRet $(c_x, p_x)$
  }
\end{minipage}
\hfill
\begin{minipage}[t]{0.45\textwidth}
  \DontPrintSemicolon
  \SetKwFunction{RLookupFunc}{$\rlookup$}
  \SetKwProg{Fn}{}{:}{}
  \Fn{\RLookupFunc{$\B$, $i$}}{
    $(n, y, \B_L, \B_R) = \B$\;
    $(n_R, n_L) = (\abs{\B_R}, \abs{\B_L})$\;
    $(c_y, p_y) = (n_L, n - (n_L + n_R))$\;
  \uIf{$c_y \leq i < c_y + p_y$}{
    $(x, c_x, p_x) = (y, c_y, p_y) $\;
  }
  \uElseIf{$i < c_y $}{
    $(c_x, p_x) = \rlookup(\B_L, i)$
  }
  \uElseIf{$i \geq c_y + p_y$}{
    $i' = i - (n - n_R)$\;
    $(c, p_x) = \rlookup(\B_R, i')$\;
    $c_x = c + (n - n_R)$\;
  }
  \KwRet $(x, c_x, p_x)$
  }
\end{minipage}
\end{algorithm*}

\

As shown in \Cref{alg:insert-remove}, all necessary operations are straightforward to implement using depth-first traversal of the BST with time complexity which scales linearly with the depth of the tree.
As long as the tree is balanced enough (which is guaranteed during encoding and highly likely during decoding), we have \(\BigO(\log m)\) complexity for each operation.
We discuss this in more detail in \ref{subsec:complexity}.
For illustrative purposes, a step through of $\mathtt{reverse\_lookup}$ is available in Appendix \Cref{app:rlookup}.
}

\subsection{Time Complexity}\label{subsec:complexity}
In practice, the dominant factors affecting the runtime of encoding and
decoding with our method are the total ($n$) and unique ($m$) number of
symbols in the multiset, as well as the complexity of encoding and decoding
with $P_X$.

The BST from \cref{subsec:swor} is used to store the elements of the multisets $\M_i$.
The BST allows lookup, insertion and removal of elements in time proportional to the depth of the tree.
Traversing the BST also requires comparing symbols under a fixed (usually lexicographic) ordering.
In the extreme case where symbols in the alphabet $\X$ are represented by binary strings of length $\log\abs{\X}$, a single comparison between symbols would require $\log\abs{\X}$ bit-wise operations.
However, `short-circuit' evaluations, where the next bits are compared only if all previous comparisons result in equality, make this exponentially unlikely.
In practice, average time may have no dependence on $\abs{\X}$.

\rebuttal{The expected and worst-case time-complexity of the multiset encoder and decoder which we propose  are shown in \Cref{tab:complexity} as a function of the cost of encoding ($E$) and decoding ($D$) with \(P_X\).}
All expected and worst-case complexities of our method are independent of the alphabet size, while current methods require at least $\BigO(\abs{\X})$ iterations.

\begin{table}[ht]
  \centering
  \caption{\rebuttal{Time complexities of our method.}}
  \begin{tabular}{ccc}\toprule
    &Encoder&Decoder\\
    \midrule
    Expected  &\(\BigO(nE + n\log m)\)&\(\BigO(nD + n\log m)\)\\
    Worst-case&\(\BigO(nE + n\log m)\)&\(\BigO(nD + nm)\)
  \end{tabular}
   \label{tab:complexity}
\end{table} 

We now detail the time dependence of sampling without replacement, and its inverse, on $n$ and $m$, assuming a fixed cost of comparing two symbols.
At the encoder, a balanced BST representing $\M_1 = \M$ must first be constructed from the sequence $x^n$.
This requires sorting followed by performing the $\bstinsert$ operation $n$ times, starting with an empty BST.
Sampling $\tilde{x}^n$ from the multiset is done via $\rlookup$ and $\bstremove$.
All operations have worst-case and average complexity equivalent to that of a search on a balanced BST, implying the overall complexity is $\BigO(n\log m)$, including the initial sorting step.

At the decoder, the BST is constructed by inserting $\tilde{x}_n, \dots, \tilde{x}_1$, which are ordered randomly, implying the BST will not always be balanced.
However, the expected depth of any node in a randomly initialized BST, which is proportional to the expected time of the \(\flookup\) and \(\bstinsert\) operations, is \(\BigO(\log m)\) \cite{knuth1998}.
The worst-case is a tree with one long, thin branch, where those operations would take \(\BigO(m)\) time.
We believe that a self-balancing tree, such as an AVL tree \cite{adelson-velsky1962} or red-black tree \cite{bayer1972}, could be used to achieve the same worst-case complexity as encoding, although we have not yet implemented those methods.

\subsection{Initial Bits}\label{subsec:initbits}
This section provides detail on the initial bits overhead of our method.
Encoding relies on alternating between a decode and encode step.
If the ANS state reaches zero at any step $i$, it must be artificially increased to allow for sampling, implying the savings at step $i$ due to decoding can be \emph{less} than $H(X_i\given\M_i)$. 
In the worst case, unlikely symbols will be sampled from the multiset early on (i.e., for small $i$), wasting potential savings.

In our experiments, this `depletion' of the integer ANS state did not seem to occur. 
There are at least two theoretical results that corroborate these findings.
First, we can easily show that for a fixed multiset $\M_i$, the expected change in message length $\Delta_i(\M_i)$ at step $i$ is always non-negative.
\changed{
At step $i$, a symbol $X_i \sim P_{X_i\given\M_i}$ is sampled via decoding and then encoded using $P_{X_i}$, resulting in
\begin{align}
    \Delta_i(\M_i)
    = \E\left[ \log P_{X_i\given\M_i}(X_i\given\M_i) - \log P_{X_i}(X_i) \right]
    = \kl{P_{X_i\given\M_i}}{P_{X_i}} \geq 0
\end{align}
Second, $X^n$ is exchangeable and therefore from 
\begin{align}
    H(X_i\given\M_i) &= H(X_i\given\M, X_{i-1}, \dots, X_2, X_1)\\
                     &\leq H(X_i\given\M, X_{i-1}, \dots, X_2)\\
                     &= H(X_{i-1}\given\M, X_{i-2}, \dots, X_1)\\
                     &= H(X_{i-1}\given \M_{i-1}),
\end{align}
we have
\begin{align}
    \rebuttal{\E}\left[\Delta_i(\M_i) - \Delta_{i-1}(\M_{i-1})\right] = -H(X_i\given\M_i) + H(X_{i-1}\given \M_{i-1}) \geq 0.
\end{align}
This suggests that the only time the state is likely to be empty is at the very beginning of encoding, i.e.,\ the initial bits overhead is a
\emph{one-time} overhead.
}

\changed{
\subsection{Analysis under Exchangeability}\label{sec:exchangeability}
As mentioned before, assuming statistical independence between symbols in $X^n$ is too strong of a condition. Exchangeability is a necessary and sufficient condition for our method to be optimal. In this sub-section we describe how to modify our algorithm and proofs to account for exchangeability. Furthermore, we show that even for non-exchangeable symbols, there exists an exchangeable distribution that can be used to optimally encode the multiset; although computing this distribution may be computationally expensive.

For exchangeable symbols, encoding $\tilde{X}_i$ must be done with the appropriate distribution that accounts for previously sampled symbols: $P_{\tilde{X}_i\given \tilde{X}^{i-1}} = P_{X_i\given X^{i-1}}$. The overall complexity of our method would still be the same as in \Cref{tab:complexity}. The caveat is that the complexity of encoding $(E)$ and decoding $(D)$ with the symbol distribution is likely to be larger than with static entropy coding (i.e., if symbols are i.i.d.). 

The rate for exchangeable symbols is
\begin{align}\label{eq:ms-rate}
    R = \sum_{i=1}^n\left( H(\tilde{X}_i\given\tilde{X}^{i-1}) - H(\tilde{X}_i\given\M_i)\right)
      = H(\M).
\end{align}
The second term remains unchanged relative to what was discussed in \Cref{subsec:rate}, as no modifications are necessary in the sampling via decoding step. It follows from \Cref{eq:exch} that $\sum_{i=1}^n H(\tilde{X}_i \given \tilde{X}^{i-1}) = H(\tilde{X}^n) = H(X^n)$ by the definition of exchangeability.

Given any distribution over sequences $P_{X^n}$, there exists an exchangeable distribution $\bar{P}_{X^n}$ that results in the same distribution over multisets $P_\M$.
Let $[x^n] = \left\{ z^n \in \X^n \colon \multiset{z^n} = \multiset{x^n} \right\}$ represent the set of sequences with the same frequency count of symbols as $x^n$.
Any sequence $z^n \in [x^n]$ creates the same multiset $\M = \multiset{z^n} = \multiset{x^n}$. 
The cardinality of $[x^n]$ is equal to the multinomial coefficient of $\M$.

The distribution over multisets induced by $P_{X^n}$ is
\begin{align}
    P_{\M}(\M) = \sum_{z^n \in [x^n]} P_{X^n}(z^n).
\end{align}
The sum is over all elements in $[x^n]$.
We can therefore exchange probability mass between sequences in $[x^n]$ without changing $P_\M(\M)$.
In particular, the distribution $\bar{P}_{X^n}$ that assigns equal mass to all sequences in the same set $[x^n]$ is, by definition, exchangeable:
\begin{align}
    \bar{P}_{X^n}(x^n) = \frac{1}{\abs{[x^n]}} P_\M(\M).
\end{align}
Therefore, for an arbitrary source distribution $P_{X^n}$, if an efficient method for computing $\bar{P}_{X^n}$ exists, our method can be used to compress $\M$ to the optimal rate.
}

\section{Experiments}\label{sec:experiments}
In this section we present experiments on synthetically generated multisets with known source distribution, multisets of grayscale images with lossy codecs, and collections of JSON maps represented as a multiset of multisets.
We used the ANS implementation in the Craystack library \cite{townsend2020a} for all experiments. 

\subsection{Synthetic multisets}
Here, synthetically generated multisets are compressed to provide evidence of the computational complexity in \cref{tab:complexity}, and optimal compression rate of the method.
We grow the alphabet size $\abs{\A}$ while sampling from the source in a way that guarantees a fixed number of unique symbols $m = 512$.
The alphabet $\A$ is always a subset of $\mathbb{N}$.
For each run, we generate a multiset with $m=512$ unique symbols, and use a skewed distribution, sampled from a Dirichlet prior with coefficients $\alpha_k = k$ for $k=1,\ldots,\abs{\A}$, as the distribution $D$.

\begin{figure}[h]
    \begin{center}
    \includegraphics[width=0.45\textwidth]{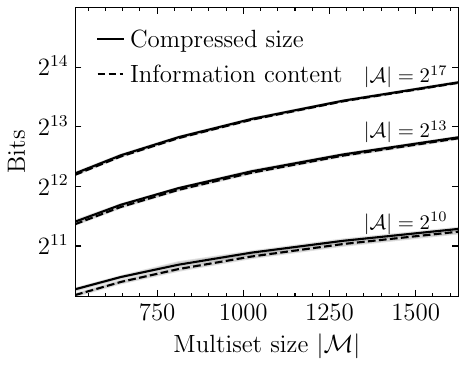}
    \hfill
    \includegraphics[width=0.45\textwidth]{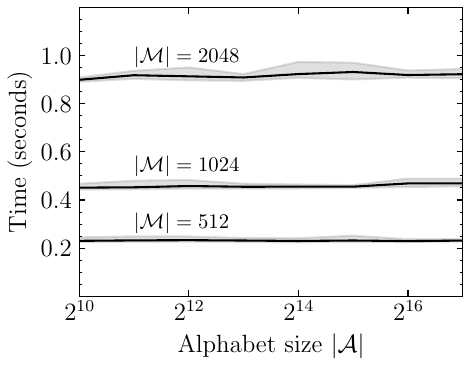}
    \end{center}
    \caption{ Left: final compressed length is close to the information content for varying alphabet and multiset sizes. Right: computational complexity does not scale with alphabet size $|\A|$, and is linear in $|\M|$.}
    \label{fig:toyms}
\end{figure}

The final compressed size of the multiset and the information content assuming \(X \sim P_X\), are shown in \Cref{fig:toyms} for different settings of \(\abs{\A}\), alongside the total encode plus decode time. Results are averaged over 20 runs, with shaded regions representing the 99\% to 1\% confidence intervals. In general, the new codec compresses a multiset close to its information content for varying alphabet sizes, as can be seen in the top plot.

The total encode plus decode time is unaffected by the alphabet size $\abs{\A}$.
As discussed previously, the overall complexity depends on that of coding under \(P_X\).
Here, the \(P_X\) codec does include a logarithmic time binary search over $\abs{\A}$, but this is implemented efficiently and the alphabet size can be seen to have little effect on overall time.
The total time scales linearly with the multiset size $\abs{\A}$ (bottom plot), as expected.

\subsection{MNIST with lossy WebP}
We implemented compression of multisets of grayscale images using the lossy WebP codec.
We tested on the MNIST test set, which is composed of $10,000$ distinct grayscale images of handwritten digits, each $28\times 28$ in size.
To encode the multiset, we perform the sampling procedure to select an image to compress, as usual.
The output of WebP is a prefix-free, variable-length sequence of bytes, which we encoded into the ANS state via a sequence of $\encode$ steps with a uniform distribution \rebuttal{over bytes (i.e., probability mass $1/256$ on each element).}
It is also possible (and faster) to move the WebP output directly into the lower-order bits of the ANS state \rebuttal{(see page 6 of \cite{giesen2014} for more details on this technique)}. 

\begin{figure}[h]
    \begin{center}
    \includegraphics[width=0.475\textwidth]{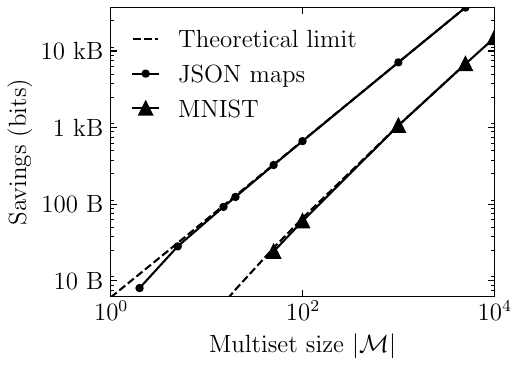}
    \hfill
    \includegraphics[width=0.45\textwidth]{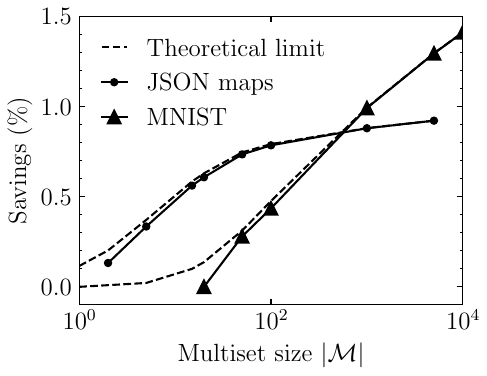}
    \end{center}
    \caption{\label{fig:lossy-mnist}%
    Rate savings due to using our method to compress a multiset instead of
  treating it as an ordered sequence. Savings are close to the theoretical
limit in both cases. The symbols are sequences of bytes outputted by lossy WebP for MNIST, and UTF-8 encoding for JSON maps. A uniform distribution over bytes is used to encode with ANS. Top: Savings in raw bits. Bottom: Percentage savings.}
\end{figure}

We compared the final compressed length with and without the sampling step. In other words, treating the dataset as a multiset and treating it as an ordered sequence. The savings achieved by using our method are shown in \Cref{fig:lossy-mnist}. The theoretical limit shown in the top plot is $\log |\M|!$, while in the bottom plot this quantity is divided by the number of bits needed to compress the data sequentially. 

Note that the maximum savings per symbol $-\frac{1}{|\M|}\log_2 |\M|! \approx
\log_2 |\M|$ depends only on the size of multiset. Therefore, when the
representation of the symbol requires a large number of bits, the percentage
savings are marginal (roughly 1.5\% for 10,000 images, in our case). To improve
percentage savings (bottom plot), one could use a better symbol codec or an adaptive codec which doesn't
treat the symbols as independent. However, as mentioned, the savings in raw bits (top plot) would remain the same, as it depends only on the multiset size $|\M|$.

\subsection{Collection of JSON maps as nested multisets}
The method can be nested to compress a multiset of multisets, by performing
an additional sampling step that first chooses which inner multiset to
compress. In this section we show results for a collection of JSON maps $\M =
\{\J_1, \dots, \J_{|\M|}\}$, where each map $\J_i = \{(k_1, v_1), \dots,
(k_{|\J_i|}, v_{|\J_i|})\}$ is itself a multiset of key-value pairs. To
compress, a depth-first approach is taken. First, some $\J \in \M$ is sampled
without replacement. Key-value pairs are then sampled from $\J$, also without
replacement, and compressed to the ANS state until $\J$ is depleted. This
procedure repeats, until the outer multiset $\M$ is empty. Assuming all maps
are unique, the maximum number of savable bits is
\begin{equation}\label{eq:nested-savings}
    \log |\M|! + \sum_{i=1}^{|\M|} \log |\J_i|!.
\end{equation}
The collection of JSON maps is composed of public GitHub user data taken from a
release of the Zstandard project\footnote{
\url{https://github.com/facebook/zstd/releases/tag/v1.1.3}}. All key-value
elements were cast to strings, for simplicity, and are encoded as UTF-8 bytes
using a uniform distribtuion. \Cref{fig:lossy-mnist}
shows the number and percentage of saved bits. The theoretical limit curve
shows the maximum savable bits with nesting, i.e.,\ \cref{eq:nested-savings}. Note that, without nesting, the theoretical limit would be that of MNIST (i.e., $\log |\M|!$).
The method gets very close to the maximum possible savings, for various numbers
of JSON maps. The rate savings were small, but these could be improved by using
a better technique to encode the UTF-8 strings.

Assuming $\J$ represents the JSON map in $\M$ with the largest number of
key-value pairs, and that the time complexity of comparing two JSON maps is
\(\BigO(|\J|)\), the complexity of the four BST operations for the \emph{outer}
multiset is \(\BigO(|\M||\J|\log|\M|)\). The overall expected time complexity
for both encoding and decoding is therefore $\BigO(|\M||\J|(\log|\M| +
\log|\J|))$.  We believe it may be possible to reduce this by performing the
inner sampling steps in parallel, or by speeding up the JSON map comparisons.

\subsection{Lossless Neural Compression on Binarized MNIST}
Originally, BB-ANS was introduced as an entropy coder for \emph{latent variable models} (LVM), such as Gaussian mixtures, that are learned from source samples (see \cite{townsend2020a} for a detailed discussion).
In this section, we use the original implementation of BB-ANS in-place of the codec for $P_X$.
The average bit-length achieved by BB-ANS is an upper-bound on the cross-entropy between the LVM and the true source distribution, and is equal to the \emph{Negative Evidence Lower Bound} \cite{townsend2020a}.
We used the pre-trained model and code made publicly available by the author\footnote{\url{https://github.com/bits-back/bits-back}}.

First, invertible sampling is performed to select a binarized MNIST image for compression.
Then, BB-ANS is applied as described in the experimental section of \cite{townsend2020a}.
This process is repeated, until all $10,000$ images are compressed.
We compared the average bit-length with and without the invertible sampling step.
Since the images are all unique, the maximum theoretical savings is $\log(10,000!) \text{ bits} \approx 14\ \text{kB}$.
This represents a potential savings of $7.6\%$, which is achieved by our method, at the cost of only $10\%$ extra computation time on average.

\section{Discussion}
As discussed in \cite{varshney2006}, a sequence can be viewed as a multiset
(frequency count of symbols) paired with a permutation (the order of symbols in
the multiset).  In this sense, our method implicitly pairs a multiset with a
permutation that is incrementally built through sampling without replacement.
Through this pairing, we convert a computationally intractable problem
(compressing a multiset) into a tractable one (compressing a sequence),
while still achieving the optimal rate. This general technique of making a
compression problem easier, by augmenting the data in some way, is known as
bits-back coding (see Related Work section). Other fruitful applications of
bits-back coding may exist which have yet to be discovered.

The stack-like nature of ANS precludes random access to symbols in the
compressed multiset, as well as streaming multiset communication, because the
full multiset must be known before encoding can begin. Methods may exist which
allow streaming and/or random access, and this may be another interesting
research direction.

We see a number of potential generalizations of the method we have presented.
Firstly, as suggested in the introduction, the independence assumption can be
relaxed: the method can also be used with an adaptive sequence compressor, as
long as the implied model over symbols is \emph{exchangeable} (that is, the
probability mass function is invariant to permutations of its inputs). This
condition is equivalent to the adaptive decoder not depending on the order of
previously observed symbols, i.e.,\ at each decoding step the symbol codec only
has access to the \emph{multiset} of symbols observed so far. 

Two examples of distributions over multisets in which symbols are exchangeable
but not i.i.d.\ are given in \cite{Steinruecken2016-oy}. The first is multisets
of symbols drawn i.i.d.\ from an \emph{unknown} distribution, where the
distribution itself is drawn from a Dirichlet prior. The unknown distribution
is effectively learned during decoding. The second is uniformly distributed
K-combinations (or submultisets) of some fixed ambient multiset. One way to
generate a uniformly distributed K-combination is to sample elements without
replacement from the ambient multiset. It is possible to efficiently compress
K-combinations using an extension of the method in this paper, taking advantage
of the fact that the ambient multiset is static and avoiding materializing it
explicitly. We leave more detailed discussion of this to future work.

As well as using the method with adaptive codecs, it would also be interesting to explore applications to more elaborate multiset-like structures.
For example, multigraphs, hypergraphs, and more sophisticated tree structured files, such as the JSON example in the experiments section, which may be represented as nested multisets all the way down.

\bibliographystyle{IEEEtran}
\bibliography{IEEEabrv, bibtex/bib/IEEEexample}

\begin{thebibliography}{10}
\providecommand{\url}[1]{#1}
\csname url@samestyle\endcsname
\providecommand{\newblock}{\relax}
\providecommand{\bibinfo}[2]{#2}
\providecommand{\BIBentrySTDinterwordspacing}{\spaceskip=0pt\relax}
\providecommand{\BIBentryALTinterwordstretchfactor}{4}
\providecommand{\BIBentryALTinterwordspacing}{\spaceskip=\fontdimen2\font plus
\BIBentryALTinterwordstretchfactor\fontdimen3\font minus
  \fontdimen4\font\relax}
\providecommand{\BIBforeignlanguage}[2]{{%
\expandafter\ifx\csname l@#1\endcsname\relax
\typeout{** WARNING: IEEEtran.bst: No hyphenation pattern has been}%
\typeout{** loaded for the language `#1'. Using the pattern for}%
\typeout{** the default language instead.}%
\else
\language=\csname l@#1\endcsname
\fi
#2}}
\providecommand{\BIBdecl}{\relax}
\BIBdecl

\bibitem{varshney2006}
L.~Varshney and V.~Goyal, ``Toward a source coding theory for sets,'' in
  \emph{2006 Data {{Compression Conference}} ({{DCC}})}.\hskip 1em plus 0.5em
  minus 0.4em\relax IEEE, 2006, pp. 13--22.

\bibitem{Steinruecken2016-oy}
C.~Steinruecken, ``Compressing {{Combinatorial Objects}},'' in \emph{2016
  {{Data Compression Conference}} ({{DCC}})}.\hskip 1em plus 0.5em minus
  0.4em\relax IEEE, 2016, pp. 389--396.

\bibitem{severo2021your}
\BIBentryALTinterwordspacing
D.~Severo, J.~Townsend, A.~J. Khisti, A.~Makhzani, and K.~Ullrich, ``Your
  dataset is a multiset and you should compress it like one,'' in \emph{NeurIPS
  2021 Workshop on Deep Generative Models and Downstream Applications}, 2021.
  [Online]. Available: \url{https://openreview.net/forum?id=vjrsNCu8Km}
\BIBentrySTDinterwordspacing

\bibitem{severo2021compressing}
D.~Severo, J.~Townsend, A.~Khisti, A.~Makhzani, and K.~Ullrich, ``Compressing
  multisets with large alphabets,'' in \emph{2022 Data Compression Conference
  (DCC)}, 2022, pp. 322--331.

\bibitem{frey1996free}
B.~J. Frey and G.~E. Hinton, ``Free energy coding,'' in \emph{Proceedings of
  Data Compression Conference-DCC'96}.\hskip 1em plus 0.5em minus 0.4em\relax
  IEEE, 1996, pp. 73--81.

\bibitem{frey1997}
B.~J. Frey, ``Bayesian networks for pattern classification, data compression,
  and channel coding,'' Ph.D. dissertation, University of Toronto, 1997.

\bibitem{townsend2019}
J.~Townsend, T.~Bird, and D.~Barber, ``Practical lossless compression with
  latent variables using bits back coding,'' in \emph{International
  {{Conference}} on {{Learning Representations}} ({{ICLR}})}, 2019.

\bibitem{duda2009}
J.~Duda, ``Asymmetric numeral systems,'' \emph{arXiv:0902.0271 [cs, math]},
  2009.

\bibitem{steinruecken2014b}
C.~Steinruecken, ``Lossless data compression,'' Ph.D. dissertation, University
  of Cambridge, 2014.

\bibitem{yang2014compression}
X.~Yang and A.~R. Barron, ``Compression and predictive distributions for large
  alphabet iid and markov models,'' in \emph{2014 IEEE International Symposium
  on Information Theory}, 2014, pp. 2504--2508.

\bibitem{yang2017minimax}
------, ``Minimax compression and large alphabet approximation through
  poissonization and tilting,'' \emph{IEEE Transactions on Information Theory},
  vol.~63, no.~5, pp. 2866--2884, 2017.

\bibitem{gripon2012compressing}
V.~Gripon, M.~Rabbat, V.~Skachek, and W.~J. Gross, ``Compressing multisets
  using tries,'' in \emph{2012 IEEE Information Theory Workshop}, 2012, pp.
  642--646.

\bibitem{Steinruecken2014-zs}
C.~Steinruecken, ``Compressing {{Sets}} and {{Multisets}} of {{Sequences}},''
  \emph{IEEE Transactions on Information Theory}, vol.~61, no.~3, pp.
  1485--1490, 2015.

\bibitem{reznik2011coding}
Y.~A. Reznik, ``Coding of sets of words,'' in \emph{2011 Data Compression
  Conference (DCC)}.\hskip 1em plus 0.5em minus 0.4em\relax IEEE, 2011, pp.
  43--52.

\bibitem{robinson1967results}
A.~H. Robinson and C.~Cherry, ``Results of a prototype television bandwidth
  compression scheme,'' \emph{Proceedings of the IEEE}, vol.~55, no.~3, pp.
  356--364, 1967.

\bibitem{Barowsky2021-wg}
M.~Barowsky, A.~Mariona, and F.~P. Calmon, ``Predictive coding for lossless
  dataset compression,'' in \emph{{IEEE} International Conference on Acoustics,
  Speech and Signal Processing ({ICASSP})}, 2021, pp. 1545--1549.

\bibitem{townsend2021lossless}
J.~Townsend, ``Lossless compression with latent variable models,'' \emph{arXiv
  preprint arXiv:2104.10544}, 2021.

\bibitem{townsend2020}
------, ``A tutorial on the range variant of asymmetric numeral systems,''
  \emph{arXiv:2001.09186 [cs, math, stat]}, 2020.

\bibitem{giesen2014}
F.~Giesen, ``Interleaved entropy coders,'' \emph{arXiv:1402.3392 [cs, math]},
  2014.

\bibitem{Kingma2019-vx}
F.~H. Kingma, P.~Abbeel, and J.~Ho, ``Bit-{{Swap}}: {{Recursive Bits}}-{{Back
  Coding}} for {{Lossless Compression}} with {{Hierarchical Latent
  Variables}},'' in \emph{International {{Conference}} on {{Machine
  Learning}}}, Oct. 2019.

\bibitem{townsend2020a}
J.~Townsend, T.~Bird, J.~Kunze, and D.~Barber, ``{{HiLLoC}}: Lossless image
  compression with hierarchical latent variable models,'' in
  \emph{International {{Conference}} on {{Learning Representations}}
  ({{ICLR}})}, 2020.

\bibitem{townsend2021a}
J.~Townsend and I.~Murray, ``Lossless compression with state space models using
  bits back coding,'' in \emph{Neural {{Compression}}: {{From Information
  Theory}} to {{Applications}} -- {{Workshop}} at {{ICLR}}}, 2021.

\bibitem{ruan2021}
Y.~Ruan, K.~Ullrich, D.~Severo, J.~Townsend, A.~Khisti, A.~Doucet, A.~Makhzani,
  and C.~J. Maddison, ``Improving {{Lossless Compression Rates}} via {{Monte
  Carlo Bits}}-{{Back Coding}},'' in \emph{International {{Conference}} on
  {{Machine Learning}}}, 2021.

\bibitem{fenwick1994}
P.~M. Fenwick, ``A new data structure for cumulative frequency tables,''
  \emph{Software: Practice and Experience}, vol.~24, no.~3, pp. 327--336, 1994.

\bibitem{moffat1999}
A.~Moffat, ``An improved data structure for cumulative probability tables,''
  \emph{Software: Practice and Experience}, vol.~29, no.~7, pp. 647--659, 1999.

\bibitem{knuth1998}
D.~E. Knuth, \emph{The Art of Computer Programming, Volume 3}.\hskip 1em plus
  0.5em minus 0.4em\relax Addison Wesley Longman Publishing Co., Inc., 1998.

\bibitem{adelson-velsky1962}
G.~{Adelson-Velsky} and E.~Landis, ``An algorithm for the organization of
  information,'' \emph{Soviet Mathematics Doklady}, vol.~3, pp. 1259--1263,
  1962.

\bibitem{bayer1972}
R.~Bayer, ``Symmetric binary {{B}}-{{Trees}}: {{Data}} structure and
  maintenance algorithms,'' \emph{Acta Informatica}, vol.~1, no.~4, pp.
  290--306, 1972.

\end{thebibliography}

\newpage

\appendices
\section{Binary search Tree for Sampling without Replacement}

\subsection{BST $\rlookup$ example}\label{app:rlookup}
For illustrative purposes, we step through a $\mathtt{reverse\_lookup}$ of the multiset in \Cref{fig:bst-1}.  This
starts with  an integer $j = s \bmod (|\M| - i + 1)$ taken from the ANS state $s$.
Searching starts at the root node, and compares $j$ to the frequency and
cumulative counts, $p_x$ and $c_x$, to decide between branching left ($j <
c_x$), right ($j \geq c_x + p_x$), or returning.  For example, at the root
node, if $s=68$ and $i=1$, then $x=\mathtt{b}$, $c_{\mathtt{b}} = 1,
p_{\mathtt{b}} = 8 - 1 - 5 = 2$; \(j = 4 \geq 1 + 2 = c_\mathtt{b} +
p_\mathtt{b}\), hence we traverse to the right branch.  Branching right
requires us to re-center to focus on the sub-interval to the right
of $\mathtt{b}$.  This can be done by continuing the traversal with $j -
(c_\mathtt{b} + p_\mathtt{b})$.  Note that, if we had branched left, this
re-centering would not be necessary.  After reaching node $\mathtt{c}$, we
return back up the tree, propagating the value of \(p_\mathtt{c}\) and summing
the size of all intervals to the left of \(\mathtt{c}\) to compute
\(c_\mathtt{c}\). Note that, if we decrement (increment) the count of each node
we visit by 1, then $\rlookup$ and $\bstremove$ ($\bstinsert$) can both be performed in
one pass over the tree. The same also holds for $\flookup$.

\rebuttal{
\section{Bitsback for Multiset Compression, Visualized}\label{sec:method-viz}
In this section we provide a sequence of images, together with a small explanation, of how BB-ANS \cite{townsend2019} and Bitsback for Multiset Compression operate. 
\subsection{Bits-back with ANS (BB-ANS)}
\begin{figure}[H]
    \centering
    \includegraphics[width=0.7\textwidth]{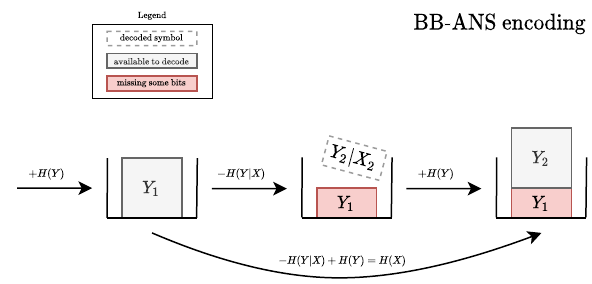}
    \caption{\rebuttal{BB-ANS encoding of i.i.d.\ symbols $X_1$ and $X_2$.
    Initially, $Y_1$ is sampled from $P_{Y\given X}(\cdot\given X_1)$ and encoded to the initially empty ANS stack using $P_Y$.
    Then, $Y_2$ is sampled from $P_{Y\given X}(\cdot\given X_2)$ using the ANS stack as a random seed, resulting in a reduction in the size of the stack.
    Finally, $Y_2$ is encoded with $P_Y$.
    This process continues until all symbols $Y_1, \dots, Y_n$ have been encoded.
    At each step the stack changes in size by $-H(Y\given X) + H(Y) = H(X)$, approaching the entropy of the source asymptotically.}}
\end{figure}

\begin{figure}[H]
    \centering
    \includegraphics[width=0.7\textwidth]{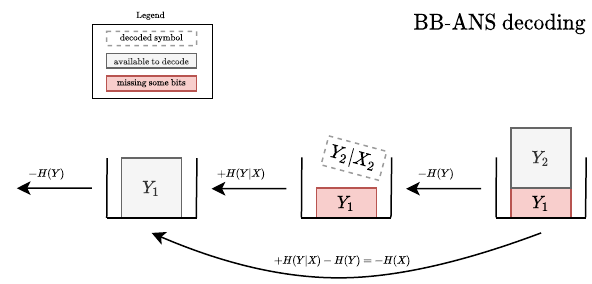}
    \caption{\rebuttal{BB-ANS decoding proceeds in the reverse order as encoding. First, $Y_2$ is decoded with $P_Y$. Then, $X_2 = f(Y_2)$ is computed and $Y_2$ is \emph{encoded} with $P_{Y \given X}(\cdot\given X_2)$. This returns to the stack the bits that were used to sample $Y_2$ during encoding, which happen to be the bits of $Y_1$. Finally, $Y_1$ is decoded with $P_Y$.}}
\end{figure}
}

\rebuttal{
\subsection{Compressing a Multiset with Bitsback}
Here we show a visualization of how our method is used to compress a multiset $\M = \{a, b, b\}$. The size of the stack $L(\M)$ is shown in bits.
\begin{figure}[H]
    \centering
    \includegraphics[width=0.8\textwidth]{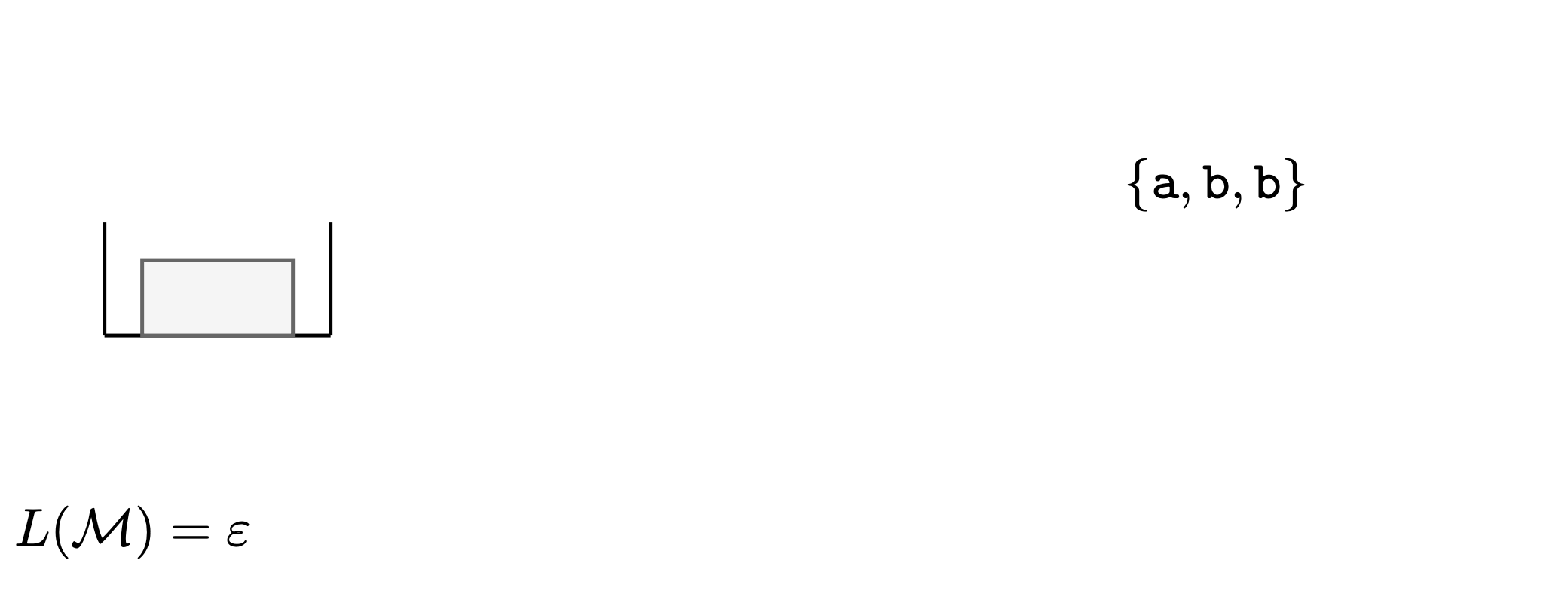}
    \caption{\rebuttal{Initially, the ANS stack is populated with some initial bits to initiate the sampling procedure.}}
\end{figure}
\begin{figure}[H]
    \centering
    \includegraphics[width=0.8\textwidth]{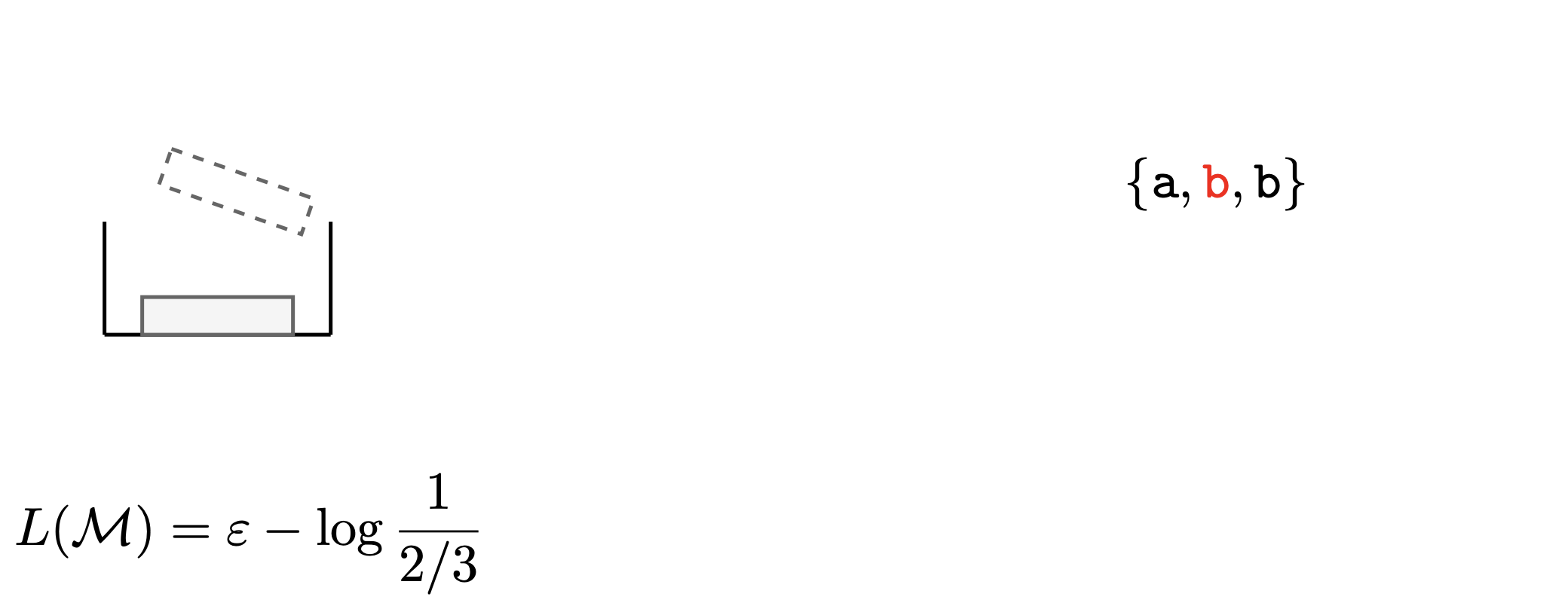}
    \caption{\rebuttal{The stack is used as a random seed to select an element in the multiset to compress. In this example, the element $\mathtt{b}$ was chosen with probability $\frac{2}{3}$, resulting in a reduction of the size of the stack.}}
\end{figure}
\begin{figure}[H]
    \centering
    \includegraphics[width=0.8\textwidth]{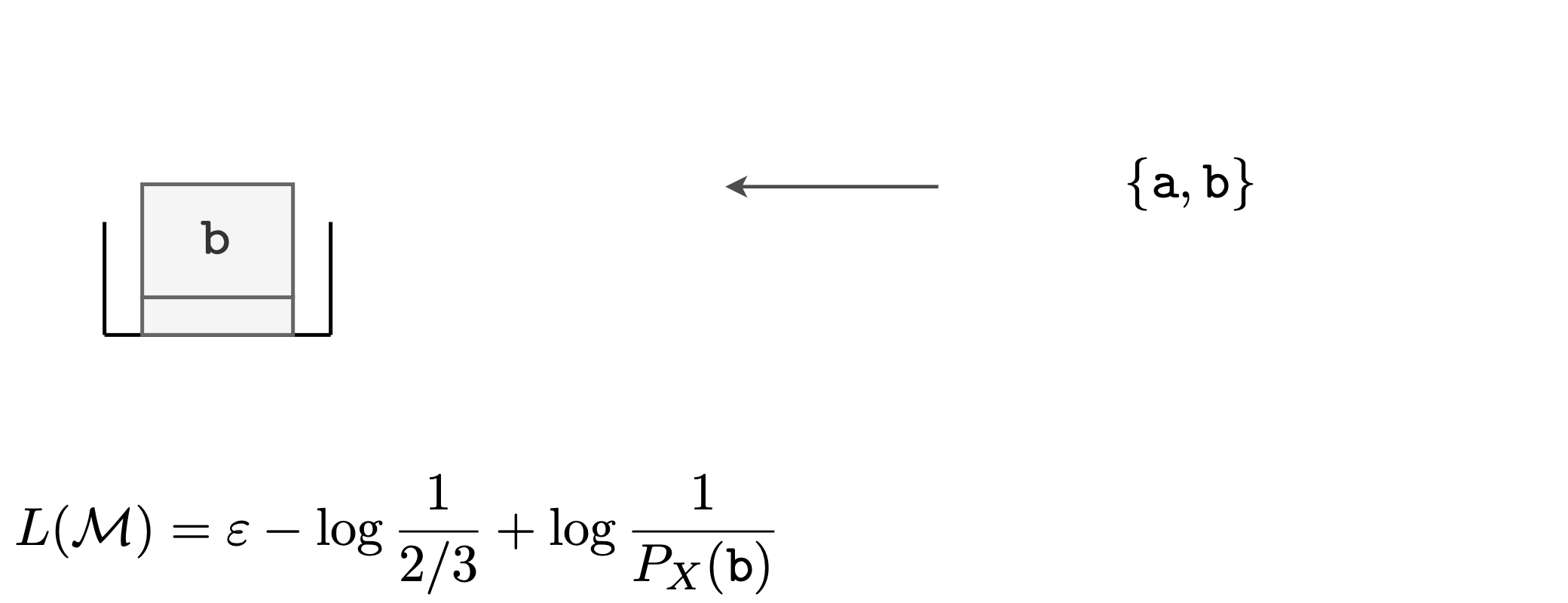}
    \caption{\rebuttal{The selected element ($\mathtt{b}$) is removed from the multiset and encoded onto the stack with $P_X(\mathtt{b})$, increasing the stack size.}}
\end{figure}
\begin{figure}[H]
    \centering
    \includegraphics[width=0.8\textwidth]{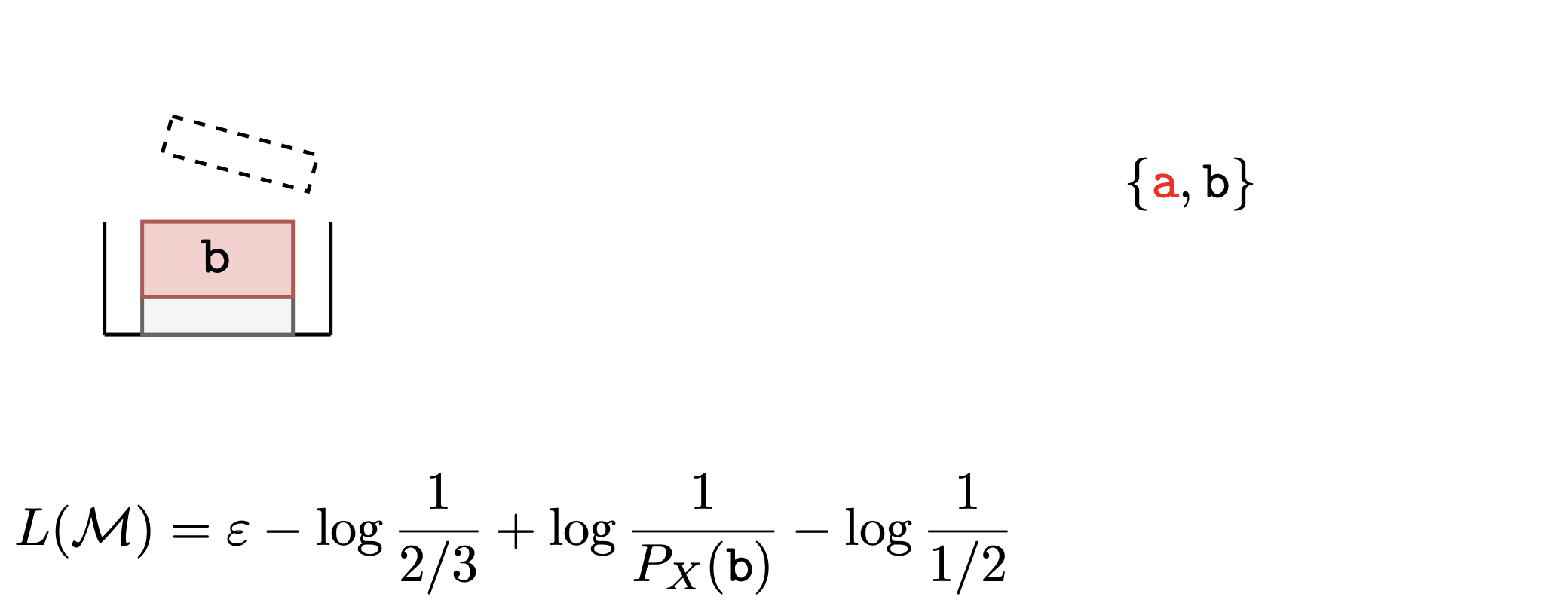}
    \caption{\rebuttal{As before, the stack is used as a random seed to select an element ($\mathtt{a}$, with probability $\frac{1}{2}$). The bits used to select $\mathtt{a}$ are from the previous encoding of $\mathtt{b}$.}}
\end{figure}
\begin{figure}[H]
    \centering
    \includegraphics[width=0.8\textwidth]{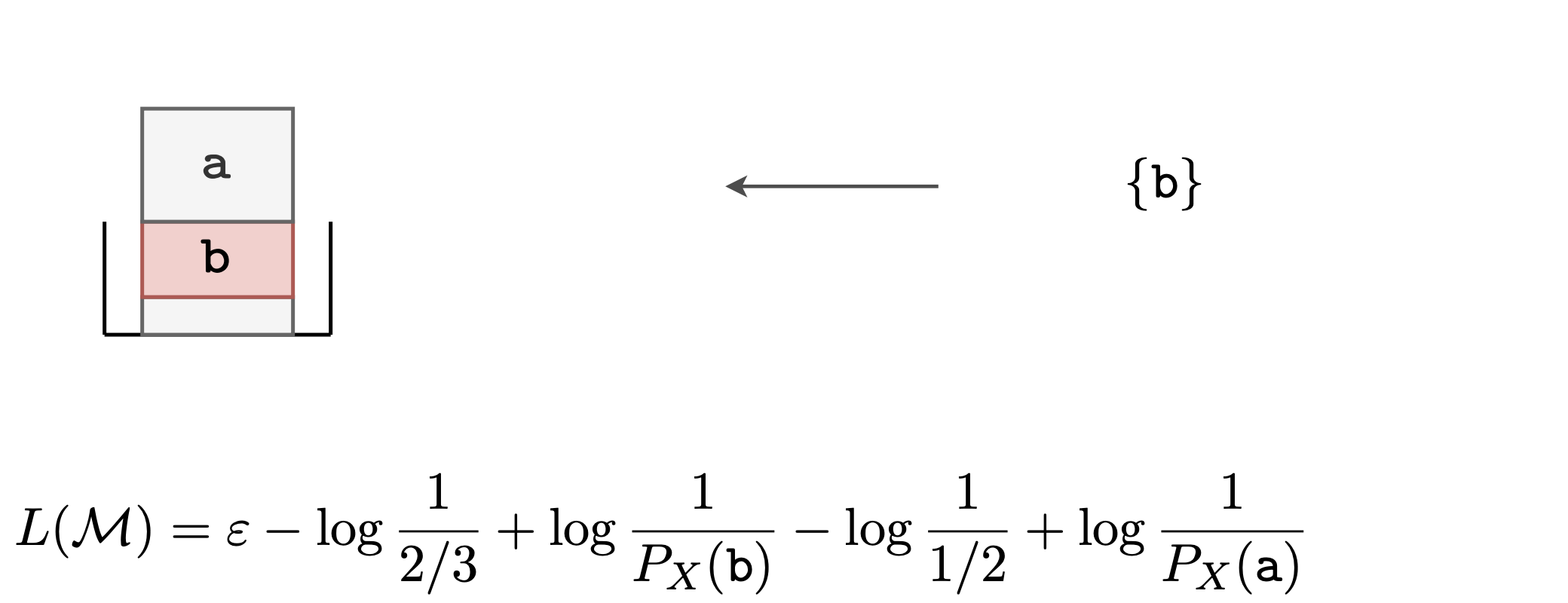}
    \caption{\rebuttal{Once again, the chosen element is encoded onto the stack.}}
\end{figure}
\begin{figure}[H]
    \centering
    \includegraphics[width=0.8\textwidth]{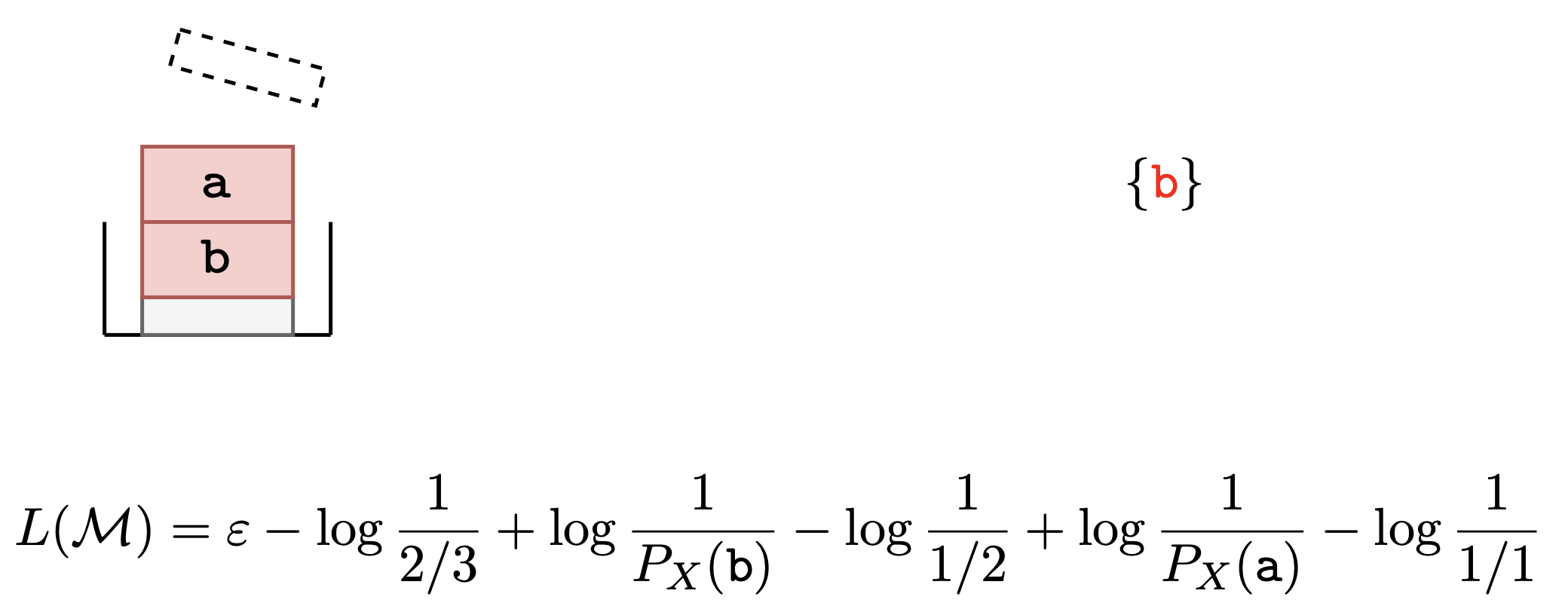}
    \caption{\rebuttal{The stack is once again used to select an element of the multiset. However, as there is only $1$ element, this operation consumes no bits.}}
\end{figure}
\begin{figure}[H]
    \centering
    \includegraphics[width=0.8\textwidth]{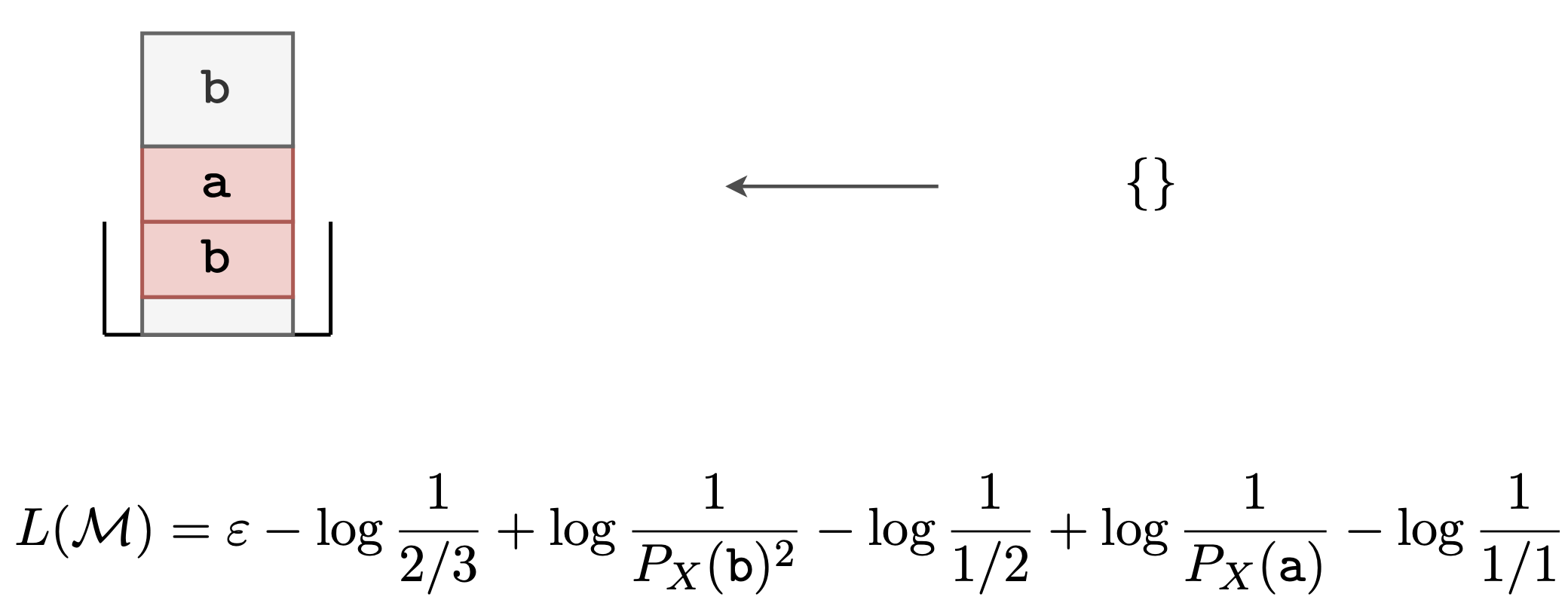}
    \caption{\rebuttal{Finally, the last element is encoded onto the stack. The total length can be shown to be $L(\M) = -\log P_\M(\{a,b,b\}) + \epsilon$.}}
\end{figure}
}

\end{document}